\documentclass[twocolumn,showpacs,preprintnumbers]{revtex4}
\usepackage{amsmath}
\usepackage{amssymb}
\usepackage{amsfonts}
\usepackage{graphicx}
\usepackage{dcolumn}
\usepackage{bm}

\input{tcilatex}

\begin{document}

\title{Quantum Games Entropy}
\author{Esteban Guevara Hidalgo$^{\dag \ddag }$}
\affiliation{$^{\dag }$Departamento de F\'{\i}sica, Escuela Polit\'{e}cnica Nacional,
Quito, Ecuador\\
$^{\ddag }$SI\'{O}N, Autopista General Rumi\~{n}ahui, Urbanizaci\'{o}n Ed%
\'{e}n del Valle, Sector 5, Calle 1 y Calle A \# 79, Quito, Ecuador}

\begin{abstract}
We propose the study of quantum games from the point of view of quantum
information theory and statistical mechanics. Every game can be described by
a density operator, the von Neumann entropy and the quantum replicator
dynamics. There exists a strong relationship between game theories,
information theories and statistical physics. The density operator and
entropy are the bonds between these theories. The analysis we propose is
based on the properties of entropy, the amount of information that a player
can obtain about his opponent and a maximum or minimum entropy criterion.
The natural trend of a physical system is to its maximum entropy state. The
minimum entropy state is a characteristic of a manipulated system i.e.
externally controlled or imposed. There exist tacit rules inside a system
that do not need to be specified or clarified and search the system
equilibrium based on the collective welfare principle. The other rules are
imposed over the system when one or many of its members violate this
principle and maximize its individual welfare at the expense of the group.
\end{abstract}

\pacs{03.65.-w, 02.50.Le, 03.67.-a, 05.30.-d}
\maketitle
\email{esteban\_guevarah@yahoo.es}

\section{Introduction}

In a recent work \cite{1} we proposed certain quantization relationships
based on the resemblances between quantum mechanics and game theory.
Although both systems analyzed are described through two apparently
different theories it was shown that both are analogous and thus exactly
equivalents. So, we can take some concepts and definitions from quantum
mechanics and physics for the best understanding of the behavior of
economical and biological processes. The quantum analogues of the replicator
dynamics and of certain relative frequencies matrix are the von Neumann
equation and the density operator respectively. This would let us analyze
the entropy of our system through the well known von Neumann (or Shannon)
entropy. The properties that these entropies enjoy would let us analyze a
\textquotedblleft game\textquotedblright\ from a different point of view
through statistical mechanics and quantum information theories.

\bigskip

There exists a strong relationship between game theories, statistical
mechanics and information theory. The bonds between these theories are the
density operator and entropy \cite{2,3}. The density operator is maybe the
most important tool in quantum mechanics. It was introduced by von Neumann
to describe a mixed ensemble in which each member has assigned a probability
of being in a determined state. From the density operator we can construct
and understand the statistical behavior about our system by using
statistical mechanics and a criterion of maximum or minimum entropy. Also we
can develop the system in function of its accessible information and analyze
it through information theories.

\bigskip

Since Shannon \cite{4,5}, information theory or the mathematical theory of
communication changed from an engineering discipline that dealed with
communication channels and codes \cite{6} to a physical theory \cite{7} in
where the introduction of the concepts of entropy and information were
indispensable to our understanding of the physics of measurement. Classical
information theory has two primary goals \cite{8}: The first is the
development of the fundamental theoretical limits on the achievable
performance when communicating a given information source over a given
communication channel using coding schemes from within a prescribed class.
The second goal is the development of coding schemes that provide
performance that is reasonably good in comparison with the optimal
performance given by the theory.

\bigskip

The von Neumann entropy \cite{2,3} is the quantum analogue of Shannon's but
it appeared 21 years before and it also generalizes Boltzmann's expression.
Entropy\textbf{\ }in quantum information theory plays prominent roles in
many contexts, e.g., in studies of the classical capacity of a quantum
channel \cite{9,10} and the compressibility of a quantum source \cite{11,12}%
. It may be defined \cite{13} as the study of the achievable limits to
information processing possible within quantum mechanics. The field of
quantum information has two tasks: First, it aims to determine limits on the
class of information processing tasks which are possible in quantum
mechanics and provides constructive means for achieving information
processing tasks. It is also the basis for a proper understanding of the
emerging fields of quantum computation \cite{14,15}, quantum communication 
\cite{16,17}, and quantum cryptography \cite{18,19}.

\bigskip

In the present work we will study the applicability of entropy and its
properties to the study of games from two sources: Information theory and
statistical mechanics and we will also propose possible applications to the
solution of specific problems in economics.

\section{On Quantum Replicator Dynamics \& the Quantization Relationships}

Evolutionary game theory combines the principles of game theory, evolution
and dynamical systems to explain the distribution of different phenotypes in
biological populations. Instead of working out the optimal strategy, the
different phenotypes in a population are associated with the basic
strategies that are shaped by trial and error by a process of natural
selection or learning. The natural selection process that determines how
populations playing specific strategies evolve is known as the replicator
dynamics \cite{20,21,22,23} whose stable fixed points are Nash equilibria 
\cite{24}.

\bigskip

We can represent the replicator dynamics in matrix commutative form and
realize that follows the same dynamic than the von Neumann equation \cite{1}%
\begin{equation}
\frac{dX}{dt}=\left[ \Lambda ,X\right] \text{.}  \label{1}
\end{equation}%
The matrix $X$ is a relative frequencies matrix. Its elements are $%
x_{ij}=\left( x_{i}x_{j}\right) ^{1/2}$ and $x_{i}$ is the relative
frequency of individuals using the strategy $s_{i}$. The matrix $\Lambda $
is equal to $\Lambda =\left[ Q,X\right] $, where $Q$ and $\Lambda $ have as
elements $q_{ii}=\frac{1}{2}\sum_{k=1}^{n}a_{ik}x_{k}$ and $(\Lambda )_{ij}=%
\frac{1}{2}\left[ \left( \sum_{k=1}^{n}a_{ik}x_{k}\right)
x_{ij}-x_{ji}\left( \sum_{k=1}^{n}a_{jk}x_{k}\right) \right] $ respectively, 
$a_{ij}$ are the elements of certain payoff matrix $A$. Each component of $X$
evolves following the replicator dynamics. If we take $\Theta =\left[
\Lambda ,X\right] $ equation (\ref{1}) becomes into $\frac{dX}{dt}=\Theta $,
where the elements of the matrix $\Theta $ are given by $\left( \Theta
\right) _{ij}=\frac{1}{2}\sum_{k=1}^{n}a_{ik}x_{k}x_{ij}+\frac{1}{2}%
\sum_{k=1}^{n}a_{jk}x_{k}x_{ji}-\sum_{k,l=1}^{n}a_{lk}x_{k}x_{l}x_{ij}$.

\bigskip

A physical or a socioeconomical system (described through quantum mechanics
and game theory respectively) is composed by $n$ members (particles,
subsystems, players, etc.). Each member is described by a state or a
strategy which has assigned a probability ($x_{i}$ or $\rho _{ij}$). The
quantum mechanical system is described by the density operator $\rho $ whose
elements represent the system average probability of being in a determined
state. In evolutionary game theory the system is described through a
relative frequencies vector $x$ whose elements represent the frequency of
players playing a determined strategy and its evolution is described through
the replicator dynamics (in matrix commutative form) which follows the same
dynamic than the evolution of the density operator (von Neumann equation).
The properties of the correspondent elements ($\rho $ and $X$) in both
systems are similar, and as expected, the properties of our quantum system
are more general than the classical system \cite{1}.

\bigskip

Although both systems are different, both are analogous and thus exactly
equivalents. This let us define and propose the next quantization
relationships%
\begin{gather}
x_{i}\longrightarrow \sum_{k=1}^{n}\left\langle i\left\vert \Psi _{k}\right.
\right\rangle p_{k}\left\langle \Psi _{k}\left\vert i\right. \right\rangle
=\rho _{ii}\text{,}  \notag \\
(x_{i}x_{j})^{1/2}\longrightarrow \sum_{k=1}^{n}\left\langle i\left\vert
\Psi _{k}\right. \right\rangle p_{k}\left\langle \Psi _{k}\left\vert
j\right. \right\rangle =\rho _{ij}\text{.}  \label{2}
\end{gather}%
A population will be represented by a quantum system in which each
subpopulation playing strategy $s_{i}$ will be represented by a pure
ensemble in the state $\left\vert \Psi _{k}(t)\right\rangle $ and with
probability $p_{k}$. The probability $x_{i}$ of playing strategy $s_{i}$ or
the relative frequency of the individuals using strategy $s_{i}$ in that
population will be represented as the probability $\rho _{ii}$ of finding
each pure ensemble in the state $\left\vert i\right\rangle $.

\bigskip

Through these quantization relationships the quantum analogue of the
replicator dynamics (in matrix commutative form) is the von Neumann equation
($i\hbar \frac{d\rho }{dt}=\left[ \hat{H},\rho \right] $), where $%
X\longrightarrow \rho $, $\Lambda \longrightarrow -\frac{i}{\hbar }\hat{H}$
and $H(x_{i})\longrightarrow S(\rho )$. These quantization relationships
could let us describe not only classical, evolutionary and quantum games but
also the biological systems that were described before through the
replicator dynamics.

\bigskip

It is important to note that equation (\ref{1}) is nonlinear while its
quantum analogue is linear i.e. the quantization eliminates the classical
system nonlinearities. The classical system that was described through the
matrix $X$ can be described now through a density operator in where its non
diagonal elements could be different from zero (and represent a mixed state)
due to the presence of coherence between quantum states which could not be
observed when the system was analyzed classically.

\section{Classical Games Entropy}

We can define the entropy of our system by%
\begin{equation}
H=-Tr\left\{ X\ln X\right\}  \label{3}
\end{equation}%
with the non diagonal elements of matrix $X$ equal to zero i.e. the Shannon
entropy over the elements of the relative frequency vector $x$%
\begin{equation}
H=-\sum_{i=1}^{n}x_{i}\ln x_{i}\text{.}  \label{4}
\end{equation}%
We can describe the evolution of the entropy of our system by supposing that
the vector of relative frequencies $x(t)$ evolves in time following the
replicator dynamics%
\begin{equation}
\frac{dx_{i}}{dt}=\left[ f_{i}(x)-\left\langle f(x)\right\rangle \right]
x_{i}=U_{i}x_{i}  \label{5}
\end{equation}%
with $U_{i}=\left[ f_{i}(x)-\left\langle f(x)\right\rangle \right] $, $%
f_{i}(x)=\sum_{j=1}^{n}a_{ij}x_{j}$ and $\left\langle f(x)\right\rangle
=\sum_{k,l=1}^{n}a_{kl}x_{k}x_{l}$%
\begin{equation}
\frac{dH}{dt}=Tr\left\{ U(\tilde{H}-X)\right\} \text{,}  \label{6}
\end{equation}%
where $\tilde{H}$ is a diagonal matrix whose trace is equal to the Shannon
entropy i.e. $H=Tr\tilde{H}$.

\section{Quantum Games Entropy}

Lets consider a system composed by $N$ members, players, strategies, etc.
This system is described completely through the density operator $\rho
(t)=\sum_{k=1}^{n}\left\vert \Psi _{k}(t)\right\rangle p_{k}\left\langle
\Psi _{k}(t)\right\vert $. Each state can \textquotedblleft
interact\textquotedblright\ with the remaining $N-1$ states. There exists a
total of $N^{2}$ states, $N$ \textquotedblleft pure\textquotedblright\ and $%
N^{2}-N$ that appear \textquotedblleft by interaction\textquotedblright\
between pure states. All of these are grouped in the matrix which represents
the density operator. The $N$ pure states are represented by the diagonal
elements of the density operator $\rho _{ii}=\sum_{k=1}^{n}\left\langle
i\left\vert \Psi _{k}\right. \right\rangle p_{k}\left\langle \Psi
_{k}\left\vert i\right. \right\rangle $ and the remaining $N^{2}-N$ states
that can appear \textquotedblleft by interaction\textquotedblright\ by the
non diagonal elements $\rho _{ij}=\sum_{k=1}^{n}\left\langle i\left\vert
\Psi _{k}\right. \right\rangle p_{k}\left\langle \Psi _{k}\left\vert
j\right. \right\rangle $.

\bigskip

In general the elements of the density operator vary in time%
\begin{gather}
i\hbar \frac{d\rho _{ii}(t)}{dt}=\sum_{l=1}^{n}(\hat{H}_{il}\rho _{li}-\rho
_{il}\hat{H}_{li})\text{,}  \notag \\
i\hbar \frac{d\rho _{ij}(t)}{dt}=\sum_{l=1}^{n}(\hat{H}_{il}\rho _{lj}-\rho
_{il}\hat{H}_{lj})\text{.}  \label{7}
\end{gather}%
When the system reaches the thermodynamic equilibrium the density operator
non diagonal elements becomes zero i.e. the coherences between stationary
states disappears while the populations of the stationary states are
exponentially decreasing functions of the energy \cite{25}.

\bigskip

We can define the entropy of our system by the von Neumann entropy%
\begin{equation}
S(t)=-Tr\left\{ \rho \ln \rho \right\}  \label{8}
\end{equation}%
which in a far from equilibrium system also vary in time until it reaches
its maximum value. When the dynamics is chaotic the variation with time of
the physical entropy goes through three successive, roughly separated stages 
\cite{26}. In the first one, $S(t)$ is dependent on the details of the
dynamical system and of the initial distribution, and no generic statement
can be made. In the second stage, $S(t)$ is a linear increasing function of
time ($\frac{dS}{dt}=const.$). In the third stage, $S(t)$ tends
asymptotically towards the constant value which characterizes equilibrium ($%
\frac{dS}{dt}=0$). With the purpose of calculating the time evolution of
entropy we approximate the logarithm of $\rho $ by series $\ln \rho =(\rho
-I)-\frac{1}{2}(\rho -I)^{2}+\frac{1}{3}(\rho -I)^{3}$... and%
\begin{eqnarray}
\frac{dS(t)}{dt} &=&\frac{11}{6}\tsum\limits_{i}\frac{d\rho _{ii}}{dt} 
\notag \\
&&-6\tsum\limits_{i,j}\rho _{ij}\frac{d\rho _{ji}}{dt}  \notag \\
&&+\frac{9}{2}\tsum\limits_{i,j,k}\rho _{ij}\rho _{jk}\frac{d\rho _{ki}}{dt}
\notag \\
&&-\frac{4}{3}\tsum\limits_{i,j,k,l}\rho _{ij}\rho _{jk}\rho _{kl}\frac{%
d\rho _{li}}{dt}+\zeta \text{.}  \label{9}
\end{eqnarray}

\section{Games Analysis from Statistical Mechanics \& QIT}

There exists a strong relationship between game theories, statistical
mechanics and information theories. The bonds between these theories are the
density operator and entropy. From the density operator we can construct and
understand the statistical behavior about our system by using statistical
mechanics. Also we can develop the system in function of its accessible
information and analyze it through information theories under a criterion of
maximum or minimum entropy.

\bigskip

Entropy \cite{4,5,27} is the central concept of information theories. The
Shannon entropy expresses the average information we expect to gain on
performing a probabilistic experiment of a random variable $A$ which takes
the values $a_{i}$ with the respective probabilities $p_{i}$. It also can be
seen as a measure of uncertainty before we learn the value of $A$. We define
the Shannon entropy of a random variable $A$ by%
\begin{equation}
H(A)\equiv H(p_{1},...,p_{n})\equiv -\sum_{i=1}^{n}p_{i}\log _{2}p_{i}\text{.%
}  \label{10}
\end{equation}%
The Shannon entropy of the probability distribution associated with the
source gives the minimal number of bits that are needed in order to store
the information produced by a source, in the sense that the produced string
can later be recovered.

\bigskip

If we define an entropy over a random variable $S^{A}$ (player's $A$
strategic space) which can take the values $s_{i}^{A}$ with the respective
probabilities $x_{i}^{A}$ i.e. $H(A)\equiv -\sum_{i=1}^{n}x_{i}\log _{2}x_{i}
$, we could interpret the entropy of our game as a measure of uncertainty
before we learn what strategy player $A$ is going to use. If we do not know
what strategy a player is going to use every strategy becomes equally
probable and our uncertainty becomes maximum and it is greater while greater
is the number of strategies. If we would know the relative frequency with
which player $A$ uses any strategy we can prepare our reply in function of
the most probable player $A$ strategy. That would be our actual best reply
which in that moment would let us maximize our payoff due to our
uncertainty. Obviously our uncertainty vanish if we are sure about the
strategy our opponent is going to use. The complete knowledge of the rules
of a game and the reserve in our strategies becomes an advantage over an
opponent who does not know the game rules or who always plays in a same
predictive way. To become a game fair, an external referee should make the
players to know completely the game rules and the strategies that the
players can use.

\bigskip

If the player $B$ decides to play strategy $s_{j}^{B}$ against player $A$
(who plays strategy $s_{i}^{A})$ our total uncertainty about the pair $(A,B)$
can be measured by an external \textquotedblleft referee\textquotedblright\
through the joint entropy of the system $H(A,B)\equiv -\sum_{i,j}x_{ij}\log
_{2}x_{ij}$, $x_{ij}$ is the joint probability to find $A$ in state $s_{i}$
and $B$ in state $s_{j}$. This is smaller or at least equal than the sum of
the uncertainty about $A$ and the uncertainty about $B,$ $H(A,B)\leq
H(A)+H(B)$. The interaction and the correlation between $A$ and $B$ reduces
the uncertainty due to the sharing of information. There can be more
predictability in the whole than in the sum of the parts. The uncertainty
decreases while more systems interact jointly creating a new only system.

\bigskip

We can measure how much information $A$ and $B$ share and have an idea of
how their strategies or states are correlated by their mutual or correlation
entropy $H(A:B)\equiv -\sum_{i,j}x_{ij}\log _{2}x_{i:j}$, with $x_{i:j}=%
\frac{\sum_{i}x_{ij}\sum_{j}x_{ij}}{x_{ij}}$. It can be seen easily as $%
H(A:B)\equiv H(A)+H(B)-H(A,B)$. The joint entropy would equal the sum of
each of $A$'s and $B$'s entropies only in the case that there are no
correlations between $A$'s and $B$'s states. In that case, the mutual
entropy vanishes and we could not make any predictions about $A$ just from
knowing something about $B$.

\bigskip

If we know that $B$ decides to play strategy $s_{j}^{B}$ we can determinate
the uncertainty about $A$ through the conditional entropy $H(A\mid B)\equiv
H(A,B)-H(B)=-\sum_{i,j}x_{ij}\log _{2}x_{i\mid j}$ with $x_{i\mid j}=\frac{%
x_{ij}}{\sum_{i}x_{ij}}$. If this uncertainty is bigger or equal to zero
then the uncertainty about the whole is smaller or at least equal than the
uncertainty about $A$, i.e. $H(A:B)\leq H(A)$. Our uncertainty about the
decisions of player $A$ knowing how $B$ and $C$ plays is smaller or at least
equal than our uncertainty about the decisions of $A$ knowing only how $B$
plays $H(A\mid B,C)\leq H(A\mid B)$ i.e. conditioning reduces entropy.

\bigskip

If the behavior of the players of a game follows a Markov chain i.e. $%
A\rightarrow B\rightarrow C$ then $H(A)\geq H(A:B)\geq H(A:C)$ i.e. the
information can only reduces in time. Also any information $C$ shares with $%
A $ must be information which $C$ also shares with $B$, $H(C:B)\geq H(C:A)$.

\bigskip

Two external observers of the same game can measure the difference in their
perceptions about certain strategy space of the same player $A$ by its
relative entropy. Each of them could define a relative frequency vector, $x$
and $y$, and the relative entropy over these two probability distributions
is a measure of its closeness $H(x\parallel y)\equiv \sum_{i}x_{i}\log
_{2}x_{i}-\sum_{i}x_{i}\log _{2}y_{i}$. We could also suppose that $A$ could
be in two possible states i.e. we know that $A$ can play of two specific but
different ways and each way has its probability distribution (again $x$ and $%
y$ that also is known). Suppose that this situation is repeated exactly $N$
times or by $N$ people. We can made certain \textquotedblleft
measure\textquotedblright , experiment or \textquotedblleft
trick\textquotedblright\ to determine which the state of the player is. The
probability that these two states can be confused is given by the classical
or the quantum Sanov's theorem \cite{6,28,29,30}.

\bigskip

By analogy with the Shannon entropies it is possible to define conditional,
mutual and relative quantum entropies which also satisfy many other
interesting properties that do not satisfy their classical analogues. For
example, the conditional entropy $S(A\mid B)$ can be negative and its
negativity always indicates that two systems (in this case players) are
entangled and indeed, how negative the conditional entropy is provides a
lower bound on how entangled the two systems are \cite{13}. If $\lambda _{i}$
are the eigenvalues of $\rho $ then von Neumann's definition can be
expressed as $S(\lambda )=-\tsum_{i}\lambda _{i}\ln \lambda _{i}$ and it
reduces to a Shannon entropy if $\rho $ is a mixed state composed of
orthogonal quantum states \cite{31}. Our uncertainty about the mixture of
states $S(\tsum_{i}p_{i}\rho _{i})$ should be higher than the average
uncertainty of the states $\tsum_{i}p_{i}S(\rho _{i})$.

\bigskip

By other hand, in statistical mechanics entropy can be regarded as a
quantitative measure of disorder. It takes its maximum possible value $\ln n$
in a completely random ensemble in which all quantum mechanical states are
equally likely and is equal to zero if $\rho $ is pure i.e. when all its
members are characterized by the same quantum mechanical state ket. Entropy
can be maximized subject to different constraints. Generally, the result is
a probability distribution function. We will maximize $S(\rho )$ subject to
the constraints $\delta Tr\left( \rho \right) =0$ and $\delta \left\langle
E\right\rangle =0$ and the result is%
\begin{equation}
\rho _{ii}=\frac{e^{-\beta E_{i}}}{\sum_{k}e^{-\beta E_{k}}}  \label{11}
\end{equation}%
which is the condition that the density operator must satisfy to our system
tends to maximize its entropy $S$. Without the internal energy constraint $%
\delta \left\langle E\right\rangle =0$ we obtain $\rho _{ii}=\frac{1}{N}$
which is the $\beta \rightarrow 0$\ limit (\textquotedblleft high -
temperature limit\textquotedblright ) in equation (\ref{11}) in where a
canonical ensemble becomes a completely random ensemble in which all energy
eigenstates are equally populated. In the opposite low - temperature limit $%
\beta \rightarrow \infty $ tell us that a canonical ensemble becomes a pure
ensemble where only the ground state is populated \cite{32}. The parameter $%
\beta $ is related inversely to the \textquotedblleft
temperature\textquotedblright\ $\tau $, $\beta =\frac{1}{\tau }$. We can
rewrite entropy in function of the partition function $Z=\sum_{k}e^{-\beta
E_{k}}$, $\beta $ and $\left\langle E\right\rangle $ via $S=\ln Z+\beta
\left\langle E\right\rangle $. From the partition function we can know some
parameters that define the system like $\left\langle E\right\rangle $ and $%
\left\langle \Delta E^{2}\right\rangle $. We can also analyze the variation
of entropy with respect to the average energy of the system%
\begin{gather}
\frac{\partial S}{\partial \left\langle E\right\rangle }=\frac{1}{\tau }%
\text{,}  \label{12} \\
\frac{\partial ^{2}S}{\partial \left\langle E\right\rangle ^{2}}=-\frac{1}{%
\tau ^{2}}\frac{\partial \tau }{\partial \left\langle E\right\rangle }
\label{13}
\end{gather}%
and with respect to the parameter $\beta $%
\begin{gather}
\frac{\partial S}{\partial \beta }=-\beta \left\langle \Delta
E^{2}\right\rangle \text{,}  \label{14} \\
\frac{\partial ^{2}S}{\partial \beta ^{2}}=\frac{\partial \left\langle
E\right\rangle }{\partial \beta }+\beta \frac{\partial ^{2}\left\langle
E\right\rangle }{\partial \beta ^{2}}\text{.}  \label{15}
\end{gather}

\section{Discussion}

In this point, it is important to remember that we are dealing with very
general and unspecific terms, definitions and concepts like state, game and
system. Due to this, the theories that have been developed around these
terms like quantum mechanics, statistical physics, information theory and
game theory enjoy of this generality quality and could be applicable to
model any system depending on what we want to mean for game, state or
system. Objectively these words can be and represent anything. Once we have
defined what system is in our model, we could try to understand what kind of
\textquotedblleft game\textquotedblright\ is developing between its members
and how they accommodate their \textquotedblleft states\textquotedblright\
in order to get their objectives. This would let us visualize what
temperature, energy and entropy would represent in our specific system
through the relationships, properties and laws that were defined before when
we described a physical system.

\bigskip

The parameter \textquotedblleft $\beta $\textquotedblright\ related with the
\textquotedblleft temperature\textquotedblright\ of a statistical system has
been used like a measure of the rationality of the players \cite{33} and in
other cases like the average amount of money per economic agent \cite{34}.
Entropy is defined over a random variable that objectively can be anything.
And depending on what the variable over which we want determinate its grade
of order or disorder is we can resolve if the best for the system is its
state of maximum or minimum entropy. If we measure the order or disorder of
our system over a resources distribution variable the best state for that
system is those in where its resources are fairly distributed over its
members which would represent a state of maximum entropy. By the other hand,
if we define an entropy over the acceptation of a presidential candidate in
a democratic process the best would represent a minimum entropy state i.e.
the acceptation of a candidate by the vast majority of the population.

\bigskip

Fundamentally, we could distinguish three states in every system: minimum
entropy, maximum entropy, and when the system is tending to whatever of
these two states. The natural trend of a physical system is to the maximum
entropy state. The minimum entropy state is a characteristic of a
\textquotedblleft manipulated\textquotedblright\ system i.e. externally
controlled or imposed. A system can be internally or externally manipulated
or controlled with the purpose of guide it to a state of maximum or minimum
entropy depending of the ambitions of the members that compose it or the
\textquotedblleft people\textquotedblright\ who control it.

\bigskip

If a physical system is not in an equilibrium state, the whole system will
vary and rearrange its state and the states of its ensembles with the
purpose of maximize its entropy which could be seen as the purpose and
maximum payoff of a physical system. The system and its members will vary
and rearrange themselves to reach the best possible state for each of them
which is also the best possible state for the whole system. This can be seen
as a microscopical cooperation between quantum objects to improve their
states in order to reach or maintain the equilibrium of the system. All the
members of our quantum system will play a game in which its maximum payoff
is the welfare of the collective. We can resume the last analysis through
what we have called as \textbf{Collective Welfare Principle:}\textit{\
\textquotedblleft A system is stable only if it maximizes the welfare of the
collective above the welfare of the individual. If it is maximized the
welfare of the individual above the welfare of the collective the system
gets unstable and eventually it collapses\textquotedblright\ }\cite{1}.

\bigskip

There exist tacit rules inside a system. These rules do not need to be
specified or clarified and search the system equilibrium under a collective
welfare principle. The other \textquotedblleft
prohibitive\textquotedblright\ and \textquotedblleft
repressive\textquotedblright\ rules are imposed over the system when one or
many of its members violate the collective welfare principle and search to
maximize its individual welfare at the expense of the group. Then it is
necessary to impose regulations on the system to try to reestablish the
broken natural order. But, this \textquotedblleft order\textquotedblright\
can not be reestablished due to the constant violation of the collective
welfare principle making this world a far from equilibrium system which
oscillate in a tendency state struggling desperately against chaos, anarchy
and its collapse.

\bigskip

Is our system the best possible? Maybe we must imitate the behavior of the
most perfect system and its equilibrium concept. In a next work we will
analyze the entropy of the world by specifying distinct random variables
which let us describe the social, political and economic behavior of our
world. Are we near to a state of maximum entropy?

\bigskip

\bigskip

\bigskip

\bigskip

\section{Conclusions}

Every game can be described by a density operator, the von Neumann entropy
and the quantum replicator dynamics. There exists a strong relationship
between game theories, statistical mechanics and information theory. The
bonds between these theories are the density operator and entropy. From the
density operator we can construct and understand the statistical behavior
about our system by using statistical mechanics. Also we can develop the
system in function of its accessible information and analyze it through
information theories under a criterion of maximum or minimum entropy
depending on over what variable we have defined entropy.

\bigskip

The words system, state and game can be and represent anything. Once we have
defined what system is in our model, we could try to understand what kind of
\textquotedblleft game\textquotedblright\ is developing between its members
and how they arrange their \textquotedblleft states\textquotedblright\ in
order to get their objectives and we could realize if would be possible to
model that system through the relationships, properties and laws that were
defined before in the study of the physical system.

\bigskip

The natural trend of a physical system is to the maximum entropy state. The
minimum entropy state is a characteristic of a \textquotedblleft
manipulated\textquotedblright\ system i.e. externally controlled or imposed.
A system\ can be internally or externally manipulated or controlled with the
purpose of guide it to a state of maximum or minimum entropy depending of
the ambitions of the members that compose it or the \textquotedblleft
people\textquotedblright\ who control it.

\bigskip

There exist tacit rules inside a system that do not need to be specified or
clarified and search the system equilibrium under the collective welfare
principle. The other\ rules are imposed over the system when one or many of
its members violate this principle and maximize its individual welfare at
the expense of the group.


\begin{thebibliography}{99}
\bibitem{1} E. Guevara H., \textit{Quantum Replicator Dynamics}, Physica A 
\textbf{369/2}, 393-407 (2006).

\bibitem{2} J. von Neumann, \textit{Thermodynamik quantummechanischer
Gesamheiten}, G\"{o}tt. Nach. \textbf{1} 273-291(1927).

\bibitem{3} J. von Neumann, \textit{Mathematische Grundlagen der
Quantenmechanik} (Springer, Berlin, 1932).

\bibitem{4} E. Guevara H., \textit{Introduction to the study of entropy in
Quantum Games}, quant-ph/0604170.

\bibitem{5} C. Shannon, \textit{A mathematical theory of communication},
Bell System Tech. Jour. \textbf{27}, 379-423 (1948).

\bibitem{6} T. M. Cover and J. A. Thomas, \textit{Elements of Information
Theory} (Wiley, New York, 1991).

\bibitem{7} R. Landauer, \textit{Information is physical}, Phys. Today 44,
23-29 (1991).

\bibitem{8} R. Gray, \textit{Entropy and Information Theory} (
Springer-Verlag, New York, 1990).

\bibitem{9} B. Schumacher and M. D. Westmoreland, Phys. Rev. A \textbf{56},
131 (1997).

\bibitem{10} A. S. Holevo, IEEE Trans. Inf. Theory \textbf{44}, 269 (1998).

\bibitem{11} B. Schumacher, Phys. Rev. A \textbf{51}, 2738 (1995).

\bibitem{12} R. Jozsa and B. Schumacher, J. Mod. Opt. \textbf{41}, 2343
(1994).

\bibitem{13} M. A. Nielsen and I. L. Chuang, \textit{Quantum Computation and
Quantum Information} (Cambridge University Press, Cambridge, 2000).

\bibitem{14} C. H. Bennett and D. P. DiVincenzo, Nature \textbf{377}, 389
(1995).

\bibitem{15} D. P. DiVincenzo, Science \textbf{270}, 255 (1995).

\bibitem{16} C. H. Bennett and S. J. Wiesner, Phys. Rev. Lett. \textbf{69},
2881 (1992).

\bibitem{17} C. H. Bennett et al., Phys. Rev. Lett. \textbf{70}, 1895 (1993).

\bibitem{18} A. Ekert, Nature \textbf{358}, 14 (1992).

\bibitem{19} C. H. Bennett, G. Brassard, and N. D. Mermin, Phys. Rev. Lett. 
\textbf{68}, 557 (1992).

\bibitem{20} J. Hofbauer and K. Sigmund, \textit{Evolutionary Games and
Replicator Dynamics} (Cambridge University Press, Cambridge, UK, 1998).

\bibitem{21} J. Weibul, \textit{Evolutionary Game Theory} (MIT Press,
Cambridge, MA, 1995).

\bibitem{22} R. Cressman, \textit{The Stability Concept of Evolutionary Game
Theory: A Dynamic Approach} (Springer-Verlag, New York, 1992).

\bibitem{23} P. D. Taylor and L. B. Jonker , \textit{Evolutionary stable
strategies and game dynamics}, Mathematical Biosciences \textbf{40}, 145-156
(1978).

\bibitem{24} R. B. Myerson, \textit{Game Theory: An Analysis of Conflict}
(MIT Press, Cambridge, 1991).

\bibitem{25} C. Tanoudji and B. Diu, Lao, \textit{Quantum Mechanics} (
Herman, Par\i s, 1977).

\bibitem{26} M. Baranger, V. Latora and A. Rapisarda, \textit{Time evolution
of thermodynamic entropy for conservative and dissipative chaotic maps},
cond-mat/0007302.

\bibitem{27} A. Wehrl, \textit{General properties of entropy}, Rev. Mod.
Phys. \textbf{50}, 221--260 (1978).

\bibitem{28} B. Schumacher and M. Westmoreland, \textit{Relative Entropy in
Quantum Information Theory}, quant-ph/0004045.

\bibitem{29} F. Hiai and D. Petz, Comm. Math. Phys. \textbf{143}, 99 (1991).

\bibitem{30} V. Vedral, M. B. Plenio, K. Jacobs, and P. L. Knight, Phys.
Rev. A \textbf{56}, 4452 (1997).

\bibitem{31} N. J. Cerf and C. Adami, Phys. Rev. Lett. \textbf{79}, 5194
(1997).

\bibitem{32} J. J. Sakurai, \textit{Modern Quantum Mechanics} (Addison -
Wesley, 1994).

\bibitem{33} R. McKelvey and T. Palfrey, \textit{Quantal Response Equilibria
for Extensive Form Games}, Experimental Economics \textbf{1}, 9-41 (1998).

\bibitem{34} A. Dr\u{a}gulescu and V. M. Yakovenko, \textit{Statistical
mechanics of money}, Eur. Phys. J. B \textbf{17}, 723-729 (2000).
\end{thebibliography}
\end{document}